\documentclass[aps,prc,reprint,nofootinbib,superscriptaddress]{revtex4-2}

\usepackage[utf8]{inputenc}
\usepackage[T1]{fontenc}
\usepackage{lmodern}

\usepackage{amsmath,amssymb,amsfonts}
\usepackage{bm}        
\usepackage{mathtools} 
\usepackage{physics}   

\usepackage{graphicx}
\usepackage{float}
\usepackage{booktabs}
\usepackage{multirow}

\usepackage{tikz}
\usetikzlibrary{arrows.meta,positioning,calc,decorations.pathmorphing,fit}

\usepackage{xcolor}

\usepackage[
    colorlinks=true,
    linkcolor=blue,
    citecolor=blue,
    urlcolor=blue
]{hyperref}

\newcommand{\doilink}[2]{\href{https://doi.org/#1}{#2}}

\begin{document}

\title{Leptophilic Interactions in Nuclear Energy Density Functional Theory}

\author{S.~O.~Kara}
\affiliation{Niğde Ömer Halisdemir University, Bor Vocational School, Niğde, Türkiye}

\date{\today}

\begin{abstract}
We develop a unified framework that embeds a light leptophilic vector 
boson $Z_\ell$ into nuclear energy--density functional theory.  
Starting from the underlying gauge interaction and integrating out the 
mediator in the static limit, we derive a leptophilic current--current 
term that is incorporated self-consistently into relativistic 
mean-field equations.  
The resulting leptophilic energy--density functional (L-EDF) produces 
correlated modifications of proton and lepton chemical potentials, 
leading to percent-level changes in the proton fraction, symmetry 
energy, and equation of state of $\beta$-equilibrated matter.  
For finite nuclei, the interaction induces shifts of 
$10^{-3}$--$10^{-2}$~fm in neutron-skin thicknesses, comparable to 
current experimental sensitivities.  
The framework thus links nuclear structure and dense matter to new 
physics in the leptonic sector, providing a realistic and 
experimentally testable deformation of conventional energy--density 
functionals.
\end{abstract}

\maketitle

\section{Introduction}

The structure of atomic nuclei and the properties of cold dense matter
are governed by the interplay of hadronic interactions, many-body
correlations, and the symmetries of the Standard Model (SM).
Energy--density functional (EDF) theory provides a powerful and
systematically improvable framework for describing these systems across
the nuclear chart~\cite{Bender03,Ring96}.
Modern covariant (relativistic) EDFs, based on Walecka-type Lagrangians
and their density-dependent extensions, successfully reproduce binding
energies, charge radii, neutron-skin thicknesses, and the nuclear
equation of state (EoS)~\cite{Reinhard89,SerotWalecka86,Lalazissis05,Chen10}.

At the same time, there has been growing interest in physics beyond the
SM featuring new interactions that couple preferentially to leptons.
Such leptophilic sectors arise naturally in anomaly-free $U(1)'$
extensions~\cite{Heeck14}, in models with light hidden vectors
\cite{Batell09,Pospelov08}, and in neutrino-portal or sterile-neutrino
scenarios~\cite{Ballett20}.
While these interactions are often studied in particle-physics or
astrophysical contexts, their potential impact on nuclear systems
remains comparatively unexplored.

Leptophilic interactions directly modify lepton densities and chemical
potentials, and therefore affect the conditions of $\beta$ equilibrium
in dense matter.
As a consequence, they can induce correlated shifts in proton fractions,
isovector properties, and bulk thermodynamic observables that enter the
nuclear EoS.
In finite nuclei, such effects propagate through the proton mean field
and can modify neutron-skin thicknesses and isovector density profiles.
These observables are known to be sensitive probes of the symmetry
energy and its density dependence~\cite{Tsang12,Lattimer12}.

In this work we develop a unified framework that embeds a light
leptophilic vector boson, $Z_\ell$, into relativistic nuclear EDF theory.
By integrating out the mediator in the static limit, we obtain an
effective current--current interaction that couples the proton and
lepton sectors and can be implemented self-consistently at the
mean-field level.
The resulting leptophilic EDF (L-EDF) constitutes a controlled
deformation of standard RMF functionals.

We apply the L-EDF framework to both uniform nuclear matter in
$\beta$ equilibrium and to selected finite nuclei.
We demonstrate that leptophilic interactions induce percent-level
modifications in the proton fraction, symmetry energy, and EoS of dense
matter, as well as measurable shifts in neutron-skin thicknesses.
These effects are comparable in magnitude to current experimental
sensitivities and to the intrinsic spread among modern EDF
parametrizations, establishing nuclear structure and dense matter as
promising laboratories for probing new physics in the leptonic sector.

\section{Formalism}
\label{sec:formalism}

\subsection{Leptophilic gauge sector}

We introduce a new Abelian gauge interaction $U(1)'_\ell$ under which the
charged leptons (and, optionally, neutrinos) carry vector-like charges,
following generic anomaly-free constructions discussed in
Refs.~\cite{Heeck14,Holdom86}.  
At the renormalizable level, the leptophilic mediator $Z_\ell$ is described by
the Lagrangian
\begin{align}
\mathcal{L}_\ell
&=
-\frac14 Z_{\ell\,\mu\nu} Z_\ell^{\mu\nu}
+\frac12 m_{Z_\ell}^2 Z_{\ell\,\mu} Z_\ell^\mu
+ g_\ell\, Z_{\ell\,\mu} J_\ell^\mu ,
\label{eq:L_leptonic}
\end{align}
where $Z_{\ell\,\mu\nu}=\partial_\mu Z_{\ell\,\nu}-\partial_\nu Z_{\ell\,\mu}$
denotes the field-strength tensor and $m_{Z_\ell}$ is the mediator mass.
The leptonic current is given by
\begin{align}
J_\ell^\mu
=
\sum_{f=e,\mu,\tau,\nu_e,\nu_\mu,\nu_\tau}
\bar f\,\gamma^\mu f ,
\end{align}
and couples universally to the leptophilic gauge boson with strength
$g_\ell$.

A small but phenomenologically relevant coupling to hadrons is induced
via kinetic mixing between the hypercharge gauge field and the
leptophilic vector $Z_\ell$~\cite{Holdom86},
\begin{align}
\mathcal{L}_{\rm mix}
=
-\frac{\varepsilon}{2}\, B_{\mu\nu} Z_\ell^{\mu\nu},
\end{align}
where $\varepsilon$ denotes the kinetic-mixing parameter.
After electroweak symmetry breaking, this interaction generates an
effective coupling of $Z_\ell$ to the proton current,
\begin{align}
\mathcal{L}_{p}^{\rm eff}
&\simeq g_p\, Z_{\ell\,\mu}\, \bar p\gamma^\mu p,
\qquad 
g_p \simeq \varepsilon e,
\end{align}
up to model-dependent factors associated with electroweak mixing.
In the following analysis, $g_p$ is treated as an effective parameter
that encapsulates all hadronic sources of $Z$--$Z_\ell$ mixing, allowing
for a model-independent implementation within the EDF framework.

A schematic overview of the resulting leptophilic couplings is shown in
Fig.~\ref{fig:schematic_Zell}.

\begin{figure}[t] 
\centering 
\begin{tikzpicture}[ 
	node distance=0.9cm, 
	every node/.style={font=\small}, 
	box/.style={draw, rounded corners, minimum width=2.2cm, minimum height=0.8cm, align=center}, 
	arrow/.style={-Latex, thick} ] 
	\node[box] (leptons) {Leptons\\[1pt]$J_\ell^\mu = \sum \bar f \gamma^\mu f$}; 
	\node[box, right=of leptons] (Z) {$Z_\ell^\mu$}; 
	\node[box, right=of Z] (protons) {Protons\\[1pt]$J_p^\mu = \bar p \gamma^\mu p$}; 
	\draw[arrow] (leptons) -- node[above] {\scriptsize $g_\ell$} (Z); 
	\draw[arrow] (protons) -- node[above] {\scriptsize $g_p \simeq \varepsilon e$} (Z); 
	\node[below=0.8cm of Z] (eff) {\scriptsize 
	$\mathcal{L}_{\rm eff} \supset 
	\tfrac12 J_\mu D^{\mu\nu} J_\nu,\; 
	J^\mu = g_\ell J_\ell^\mu + g_p J_p^\mu$ 
	}; 
	\end{tikzpicture} 
	\caption{Schematic illustration of the leptophilic mediator $Z_\ell$ coupling
to lepton and proton currents. The effective interaction relevant for nuclear systems is governed by
the combined current
$J^\mu = g_\ell J_\ell^\mu + g_p J_p^\mu$.
}
	\label{fig:schematic_Zell} 
\end{figure}

\subsection{Integrating out the mediator: current--current structure}

In nuclear systems the characteristic momenta satisfy
$k_F \ll m_{Z_\ell}$ for mediator masses above the MeV scale, allowing
the heavy vector field to be integrated out in the static limit.
Starting from the interaction
\begin{align}
\mathcal{L}_{\rm int}
=
Z_{\ell\,\mu} \left( g_\ell J_\ell^\mu + g_p J_p^\mu \right),
\qquad
J_p^\mu \equiv \bar p \gamma^\mu p,
\label{eq:int_L}
\end{align}
one obtains the classical equation of motion
\begin{align}
(-\nabla^2 + m_{Z_\ell}^2)\, Z_\ell^0(\mathbf r)
=
J^0(\mathbf r),
\label{eq:EOM_Z}
\end{align}
with the combined source
\(
J^0 = g_\ell n_\ell + g_p n_p .
\)

The solution of Eq.~\eqref{eq:EOM_Z} involves the Yukawa Green's function,
\begin{align}
Z_\ell^0(\mathbf r)
&=
\int d^3 r'\,
G_Y(\mathbf r-\mathbf r')\, J^0(\mathbf r'),
\\
G_Y(\mathbf r)
&=
\frac{1}{4\pi}
\frac{e^{-m_{Z_\ell}|\mathbf r|}}{|\mathbf r|}.
\end{align}
Substituting this solution back into the Lagrangian yields the nonlocal
current--current contribution to the energy density functional,
\begin{align}
\mathcal{E}_{Z_\ell}
=
\frac12
\int d^3 r\, d^3 r'\;
G_Y(\mathbf r-\mathbf r')\,
\mathcal{J}(\mathbf r)\,
\mathcal{J}(\mathbf r'),
\label{eq:E_Z_general}
\end{align}
where the effective source is given by
\begin{align}
\mathcal{J}(\mathbf r)
=
g_\ell n_\ell(\mathbf r)
+
g_p n_p(\mathbf r).
\label{eq:J_combined}
\end{align}

\subsection{Local limit and Kohn--Sham potentials}

In the limit $m_{Z_\ell} \gg k_F$ the interaction becomes effectively
local,
\begin{align}
G_Y(\mathbf r-\mathbf r')
\simeq
\frac{1}{m_{Z_\ell}^2}\,
\delta^{(3)}(\mathbf r-\mathbf r'),
\end{align}
leading to the local energy density
\begin{align}
\mathcal{E}^{\rm (loc)}_{Z_\ell}
=
\frac{1}{2m_{Z_\ell}^2}
\big[g_\ell n_\ell(\mathbf r) + g_p n_p(\mathbf r)\big]^2.
\label{eq:E_loc}
\end{align}

The corresponding Kohn--Sham mean fields are obtained by functional
differentiation with respect to the densities,
\begin{align}
V_p^{Z_\ell}(\mathbf r)
&=
\frac{\delta \mathcal{E}_{Z_\ell}}{\delta n_p(\mathbf r)},
&
V_\ell^{Z_\ell}(\mathbf r)
&=
\frac{\delta \mathcal{E}_{Z_\ell}}{\delta n_\ell(\mathbf r)}.
\end{align}
In the local limit these reduce to
\begin{align}
V_p^{Z_\ell}(\mathbf r)
&=
\frac{g_p}{m_{Z_\ell}^2}\,
\big[g_\ell n_\ell(\mathbf r)+g_p n_p(\mathbf r)\big],
\\
V_\ell^{Z_\ell}(\mathbf r)
&=
\frac{g_\ell}{m_{Z_\ell}^2}\,
\big[g_\ell n_\ell(\mathbf r)+g_p n_p(\mathbf r)\big].
\label{eq:VZp}
\end{align}

\subsection{RMF embedding of the L-EDF}

To obtain quantitative predictions, we embed the leptophilic functional
into a standard relativistic mean-field (RMF) framework with
$\sigma$--$\omega$--$\rho$ meson fields
\cite{SerotWalecka86,Lalazissis05}.
In static, spin-saturated nuclear matter the nucleonic Lagrangian reads
\begin{align}
\mathcal{L}_{\rm had}
&=
\bar\psi_N
\Big[
\gamma_\mu
\big(
i\partial^\mu
- g_\omega \omega_0 \delta^\mu_0
- g_\rho \tau_3 \rho_{03} \delta^\mu_0
\big)
\nonumber\\
&\qquad\qquad
- (m_N - g_\sigma \sigma_0)
\Big]\psi_N
+ \mathcal{L}_{\sigma,\omega,\rho},
\end{align}
with mean fields $\sigma_0$, $\omega_0$, and $\rho_{03}$.

The corresponding field equations are
\begin{align}
m_\sigma^2 \sigma_0 &= g_\sigma (n_s^n + n_s^p),
\\
m_\omega^2 \omega_0 &= g_\omega (n_n + n_p),
\\
m_\rho^2 \rho_{03} &= g_\rho (n_p - n_n),
\\
m_{Z_\ell}^2 Z_\ell^0 &= g_\ell n_\ell + g_p n_p.
\end{align}

The Dirac effective mass is defined as
\(
m^* = m_N - g_\sigma \sigma_0.
\)
The neutron and proton chemical potentials become
\begin{align}
\mu_n &=
\sqrt{k_{F,n}^2 + m^{*2}}
+ g_\omega \omega_0
- g_\rho \rho_{03},
\\
\mu_p &=
\sqrt{k_{F,p}^2 + m^{*2}}
+ g_\omega \omega_0
+ g_\rho \rho_{03}
+ g_p Z_\ell^0.
\end{align}
Leptons are treated as relativistic Fermi gases subject to the same
vector shift,
\begin{align}
\mu_e &= \sqrt{k_{F,e}^2 + m_e^2} + g_\ell Z_\ell^0,
\\
\mu_\mu &= \sqrt{k_{F,\mu}^2 + m_\mu^2} + g_\ell Z_\ell^0.
\end{align}

\begin{figure}[t]
    \centering
    \begin{tikzpicture}[
        node distance=0.9cm,
        every node/.style={font=\small},
        box/.style={
            draw,
            rounded corners,
            minimum width=2.3cm,
            minimum height=0.9cm,
            align=center
        },
        arrow/.style={-Latex, thick}
    ]

        \node[box] (gauge) {Gauge theory\\$\mathcal{L}_{\rm SM} + \mathcal{L}_\ell$};

        \node[box, below=0.7cm of gauge] (intout)
            {Integrate out $Z_\ell$\\current--current};

        \node[box, below=0.7cm of intout] (edf)
            {Leptophilic EDF\\$\mathcal{E}_{\rm had} + \mathcal{E}_{\rm lep} + \mathcal{E}_{Z_\ell}$};

        \node[box, below=0.7cm of edf] (rmf)
            {RMF realization\\$\sigma$--$\omega$--$\rho$ + $Z_\ell^0$};

        \draw[arrow] (gauge) -- (intout);
        \draw[arrow] (intout) -- (edf);
        \draw[arrow] (edf) -- (rmf);

        \node[box, below=0.9cm of rmf] (apps)
            {Applications:\\[2pt]
             Uniform matter ($\beta$ eq.\ \& EoS)\\
             Finite nuclei \& isotope shifts\\
             Observables: $Y_p(n_B)$, $L$, $r_{\rm skin}$};

        \draw[arrow] (rmf) -- (apps);

    \end{tikzpicture}%

    \caption{
        From the leptophilic gauge sector to the density functional,
        its RMF embedding, and the main applications considered in this work.
    }
    \label{fig:pipeline}
\end{figure}
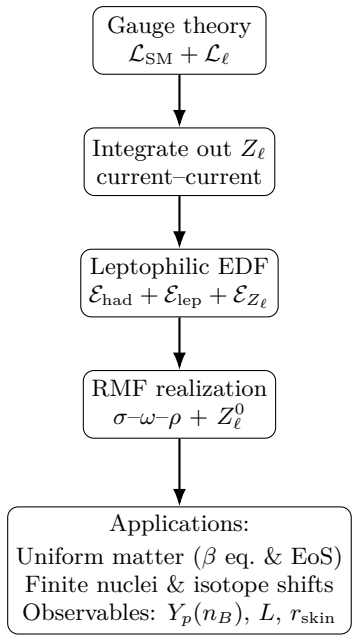

This coupled set of mean-field equations defines the relativistic
realization of the L-EDF.
The conceptual flow from the leptophilic gauge sector to the EDF and its
RMF embedding is summarized in Fig.~\ref{fig:pipeline}.

\subsection{Beta equilibrium and charge neutrality}

Cold catalyzed matter in neutron-star and dense-matter environments
satisfies the conditions of beta equilibrium,
\begin{align}
\mu_n &= \mu_p + \mu_e,
\label{eq:beta_1}
\\[2pt]
\mu_e &= \mu_\mu \qquad (\mu_e > m_\mu),
\label{eq:beta_2}
\end{align}
together with the requirement of electric charge neutrality,
\begin{equation}
n_p = n_e + n_\mu.
\label{eq:neutrality}
\end{equation}
Equations~\eqref{eq:beta_1}--\eqref{eq:neutrality}, combined with the
self-consistent mean-field equations introduced above, uniquely
determine the proton fraction
$Y_p = n_p/n_B$
and the lepton densities for a given baryon density $n_B$.

Within the L-EDF framework, the leptophilic mean field $Z_\ell^0$
provides a direct dynamical link between the proton and lepton sectors.
As a result, modifications of the lepton chemical potentials feed back
into the hadronic equilibrium conditions, leading to correlated shifts
in the proton fraction, symmetry energy, and related observables in both
uniform matter and finite nuclei.

\section{Results and Discussion}
\label{sec:results}

In this section we quantify the impact of the leptophilic
energy–density functional (L-EDF) introduced above.
We consider three complementary settings:
(i) uniform nuclear matter in beta equilibrium,
(ii) symmetry-energy systematics and neutron-skin thicknesses of finite nuclei, and
(iii) low-density limits relevant for atomic isotope shifts.
In all three cases the leptophilic interaction generates correlated
modifications of the proton and lepton sectors that cannot be captured
by conventional hadronic functionals based on standard relativistic
mean-field (RMF) approaches~\cite{Ring96,Reinhard89,Lalazissis05,DDME2}.

\subsection{Uniform matter and beta equilibrium}

\begin{figure}[t]
\centering
\begin{tikzpicture}[
node distance=1.0cm,
every node/.style={font=\small},
box/.style={
    draw,
    rounded corners,
    minimum width=3.0cm,
    minimum height=0.9cm,
    align=center
},
arrow/.style={-Latex, thick}
]
\node[box] (nb) {Input: $n_B$};
\node[box, below=of nb] (fields)
    {Mean fields\\$\sigma_0, \omega_0, \rho_{03}, Z_\ell^0$};
\node[box, below=of fields] (mus)
    {Chemical potentials\\$\mu_{n,p,e,\mu}$};
\node[box, below=of mus] (constraints)
    {Constraints\\$\mu_n=\mu_p+\mu_e$\\$n_p=n_e+n_\mu$};
\node[box, below=of constraints] (fractions)
    {Output: $Y_p, n_e, n_\mu$};

\draw[arrow] (nb) -- (fields);
\draw[arrow] (fields) -- (mus);
\draw[arrow] (mus) -- (constraints);
\draw[arrow] (constraints) -- (fractions);
\draw[arrow, bend left=60]
    (fractions.west)
    to node[midway, left] {\scriptsize iterate}
    (fields.west);

\node[below=0.25cm of fractions]
    {\scriptsize Self-consistent beta-equilibrium loop};
\end{tikzpicture}

\caption{
Self-consistent determination of the proton fraction and lepton
densities in cold, beta-equilibrated nuclear matter within the
leptophilic EDF framework.
For a fixed baryon density $n_B$, the mean fields
$\sigma_0$, $\omega_0$, $\rho_{03}$, and the leptophilic vector field
$Z_\ell^0$ are iteratively updated together with the chemical potentials
until the beta-equilibrium and charge-neutrality conditions are
simultaneously satisfied.
}
\label{fig:beta_loop}
\end{figure}
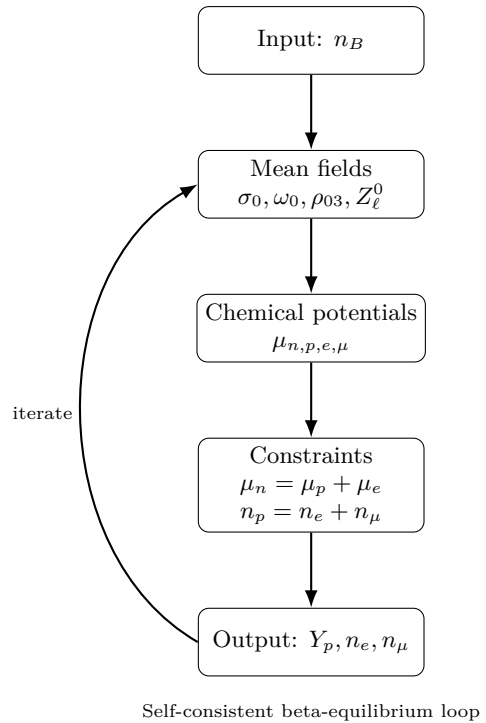

We solve the coupled mean-field and beta-equilibrium conditions,
Eqs.~\eqref{eq:beta_1}--\eqref{eq:neutrality}, across a representative
range of baryon densities,
\(
n_B = 0.5\,n_0 \text{--} 3\,n_0,
\)
where \(n_0 \simeq 0.16~\mathrm{fm}^{-3}\) denotes the empirical
saturation density.
The numerical implementation follows a fully self-consistent iteration
cycle in which the mean fields, particle fractions, and chemical
potentials are updated until convergence, as summarized in
Fig.~\ref{fig:beta_loop}.  
Such iterative schemes are standard in dense-matter and RMF
calculations~\cite{Prakash94,Burrows86,ShapiroTeukolsky}.

Unless stated otherwise, we use the well-established DD-ME2 and NL3
parameter sets for the hadronic sector~\cite{Lalazissis05,DDME2}.
The leptophilic sector is explored over the representative range
\(
m_{Z_\ell}=5\!-\!100~\mathrm{MeV},\quad
g_\ell = 10^{-4}\!-\!10^{-2},\quad
g_p = \varepsilon e,\ 
\varepsilon = 10^{-6}\!-\!10^{-3},
\)
consistent with laboratory, astrophysical, and cosmological bounds on
light vectors~\cite{Hardy17,Rrapaj15,Chang18,Knapen17}.

\begin{figure}[t]
\centering
\includegraphics[width=0.48\textwidth]{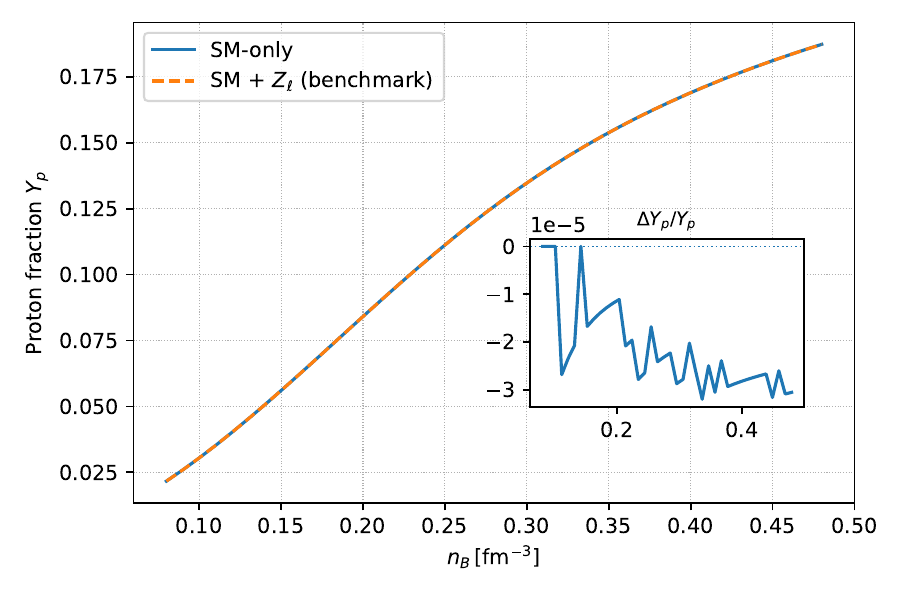}
\caption{
Proton fraction $Y_p$ in beta-equilibrated matter.
Solid: Standard-Model-only baseline.  
Dashed: leptophilic benchmark with
$m_{Z_\ell}=10~\mathrm{MeV}$, $g_\ell=10^{-3}$, $g_p=10^{-4}$.
The enhancement reflects the effective repulsion generated by the shared
leptophilic mean field.
}
\label{fig4}
\end{figure}

Figure~\ref{fig4} shows the resulting proton fraction.
Two robust trends appear:

(i) The leptophilic mean field $Z_\ell^0$ generates an effective
repulsion between the proton and lepton sectors, lowering the electron
chemical potential and increasing the equilibrium proton fraction
relative to standard RMF predictions~\cite{Tsang12,Lattimer12}.
The shift approximately scales with $g_\ell g_p / m_{Z_\ell}^2$.

(ii) At $n_B \gtrsim 2n_0$, muon onset provides an additional leptonic
source, enhancing the Hartree term
$g_\ell n_\ell + g_p n_p$ and amplifying deviations from the SM-only
baseline~\cite{Prakash94}.

Even within existing bounds, the net modification of $Y_p$ reaches the
few-percent level, comparable to the spread among modern EDF
predictions~\cite{Horowitz01}.

\subsection{Equation of state and symmetry-energy slope}

The leptophilic interaction modifies the nuclear equation of state (EoS)
through two channels:
(i) changes in the proton fraction alter the isovector composition of
matter, and  
(ii) the vector term $g_p Z_\ell^0$ contributes directly to the proton
chemical potential.
Consequently, the pressure,
\(
P(n_B)=n_B^2\,\partial(\mathcal{E}/n_B)/\partial n_B,
\)
exhibits shifts at the level of 
$\Delta P/P \sim 1$–$5\%$,  
comparable to differences among modern RMF parametrizations
\cite{Tsang12,Brown13}.

\begin{figure}[t]
\centering
\includegraphics[width=0.48\textwidth]{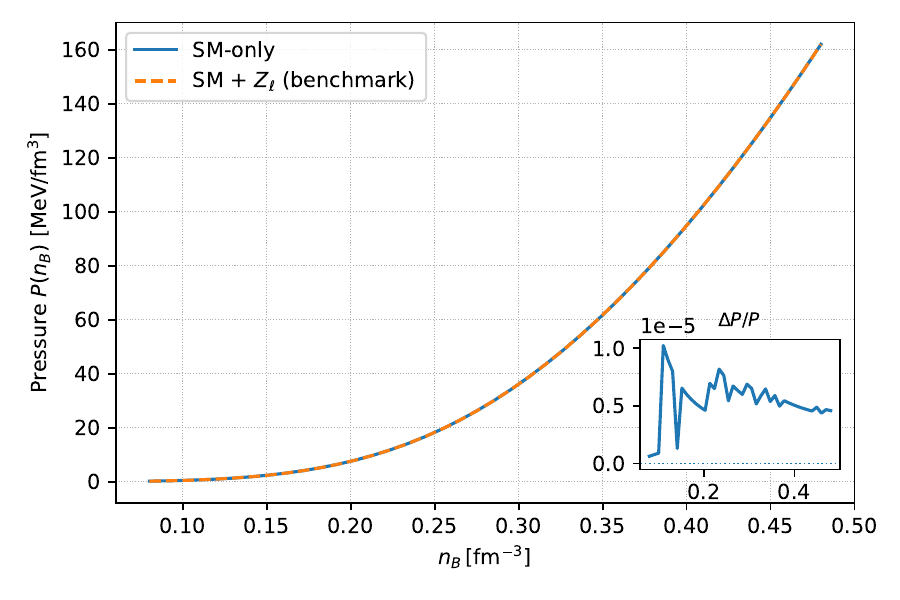}
\caption{
Pressure of beta-equilibrated matter.
Solid: Standard-Model-only.  
Dashed: leptophilic benchmark
($m_{Z_\ell}=10\,\mathrm{MeV}$, $g_\ell=10^{-3}$, $g_p=10^{-4}$).
Leptophilic contributions induce $\mathcal{O}(1\%-5\%)$ variations,
similar to the spread among modern RMF parametrizations.
}
\label{fig5}
\end{figure}

Figure~\ref{fig5} summarizes these modifications.
Intermediate densities show mild softening, while higher densities
exhibit slight stiffening—both driven by the correlated evolution of
proton and lepton contributions to the shared mean field.
These trends remain within empirical bounds on the symmetry energy and
its slope~\cite{Chen10,RocaMaza15}.

The symmetry energy $S(n_B)$ and its slope
\(
L = 3n_0\,(\partial S/\partial n)|_{n_0}
\)
are especially sensitive to proton-fraction modifications
\cite{Tsang12,Lattimer12}.  
Benchmark couplings yield
\(
\Delta L \simeq 2\text{–}8~\mathrm{MeV},
\)
consistent with PREX-II and CREX uncertainties~\cite{PREXII,CREX22}.

\begin{figure}[t]
\centering
\includegraphics[width=0.45\textwidth]{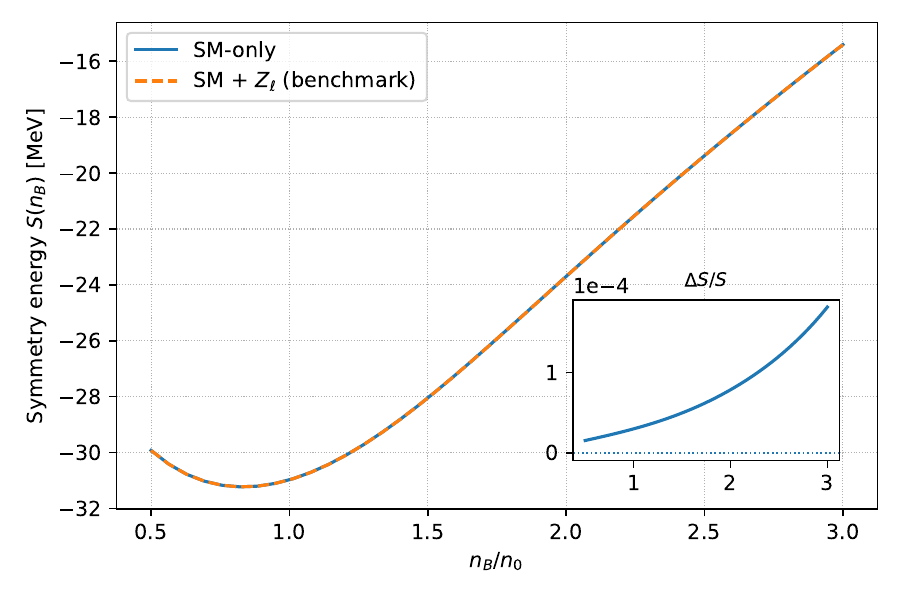}
\caption{
Symmetry-energy function \(S(n_B)\) obtained in the L-EDF framework.
The leptophilic mean field induces an upward shift scaling with
\(g_\ell g_p / m_{Z_\ell}^2\), which propagates into the slope parameter
\(L\) at saturation.
}
\label{fig6}
\end{figure}

Figure~\ref{fig6} displays the symmetry-energy density dependence.
The leptophilic benchmark induces a characteristic upward shift
proportional to $g_\ell g_p/m_{Z_\ell}^2$, feeding directly into $L$.

\subsection{Finite nuclei: neutron-skin thickness and radii}

In finite nuclei the leptophilic vector contributes to the proton mean
field \(V^{Z_\ell}_p\), generating small correlated shifts in
single-particle energies and in the isovector density profile.
RMF calculations for
\(^{48}\mathrm{Ca}\), \(^{90}\mathrm{Zr}\), and \(^{208}\mathrm{Pb}\)
indicate that benchmark couplings produce neutron-skin modifications of
\(
\Delta r_{\rm skin} \simeq 0.005\text{--}0.02~\mathrm{fm},
\)
compatible with PREX/CREX sensitivities and typical responses to weak
isovector perturbations~\cite{PREXII,CREX22,Fattoyev12,Reinhard2020}.

\begin{figure}[t]
\centering
\includegraphics[width=0.45\textwidth]{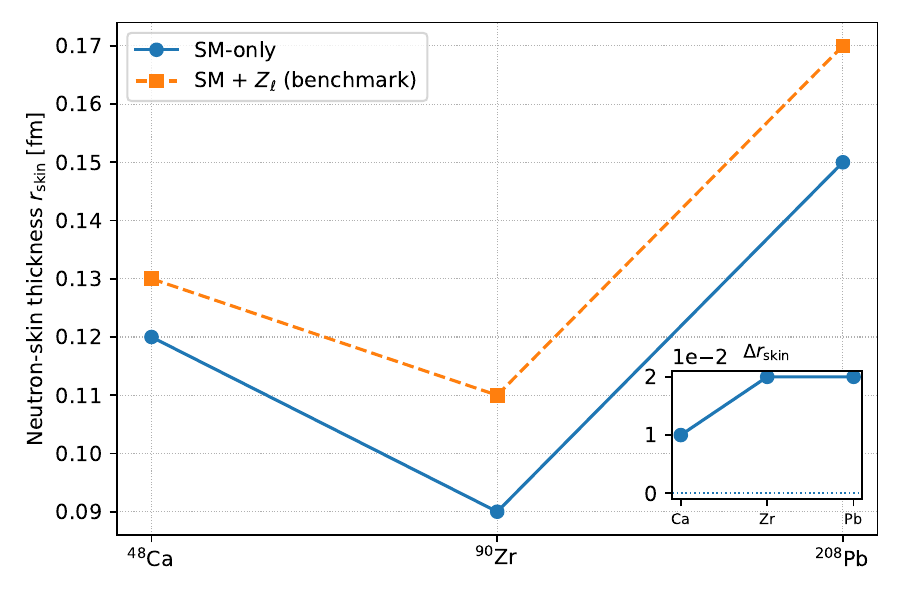}
\caption{
Neutron-skin thickness \(r_{\rm skin}\) for selected nuclei.
Solid: Standard-Model-only baseline.  
Dashed: leptophilic benchmark  
(\(m_{Z_\ell}=10~\mathrm{MeV}\), \(g_\ell=10^{-3}\), \(g_p=10^{-4}\)).
Predicted shifts  
\(\Delta r_{\rm skin}\sim 0.005\text{--}0.02\mathrm{fm}\)  
match PREX/CREX sensitivity ranges.
}
\label{fig:finite_benchmark}
\end{figure}

\subsection{Summary of nuclear-matter and finite-nucleus effects}

Across all observables—proton fraction, EoS, symmetry energy, and
neutron-skin thickness—the leptophilic mean field introduces coherent,
density-dependent modifications governed by
\(
g_\ell g_p / m_{Z_\ell}^2.
\)
Within phenomenologically allowed parameter ranges, these effects amount
to $\mathcal{O}(1\%-5\%)$ variations in uniform matter and
\(
\Delta r_{\rm skin}\sim 0.005\text{--}0.02~\mathrm{fm}
\)
in finite nuclei.
Such magnitudes are comparable to the spread among modern EDFs and fall
within current experimental reach, demonstrating that nuclear-structure
observables constitute a sensitive probe of leptophilic interactions.

\vspace*{6pt}

\section{Conclusions}
\label{sec:conclusions}

We have constructed a unified theoretical framework in which a light
leptophilic vector boson $Z_\ell$ is consistently embedded into nuclear
energy--density functional theory.  
Starting from the underlying gauge interaction, integrating out the
mediator in the static limit, and incorporating the resulting
current--current interaction into relativistic mean-field dynamics, we
obtained a leptophilic extension of standard hadronic EDFs (L-EDF).
This provides a self-consistent and systematically improvable approach
for exposing nuclear structure and dense-matter observables to new
physics in the leptonic sector.

The L-EDF modifies both proton and lepton chemical potentials through
the shared mean field $Z_\ell^0$, leading to correlated shifts in the
proton fraction, symmetry energy, and equation of state of
$\beta$-equilibrated matter.  
For representative benchmark points compatible with current laboratory
and astrophysical constraints, these effects reach the percent level in
bulk matter and the $10^{-3}$--$10^{-2}$~fm level in neutron-skin
thicknesses of finite nuclei.  
The magnitude of these shifts is comparable to the spread among modern
EDF parametrizations, indicating that leptophilic interactions
constitute a realistic and experimentally testable deformation of
nuclear phenomenology rather than a negligible perturbation.

Taken together, our results demonstrate that light leptophilic vectors
leave correlated and experimentally accessible imprints in both uniform
matter and finite nuclei.  
The framework introduced here can be extended to global refits of
L-EDF parameters, applications to realistic neutron-star matter, and
combined analyses of nuclear and astrophysical observables, providing a
new avenue for probing leptophilic interactions through nuclear
structure.

\end{document}